\documentclass[twocolumn,english,showpacs,preprintnumbers,amsmath,amssymb,floatfix]{revtex4}
\usepackage[T1]{fontenc}
\usepackage{color}
\usepackage{array}
\usepackage{amstext}
\usepackage{graphicx}
\usepackage{esint}
\usepackage[colorlinks = true,
linkcolor = magenta,
urlcolor  = blue,
citecolor = red,
anchorcolor = blue]{hyperref}

\makeatletter

\@ifundefined{textcolor}{}
{%
	\definecolor{BLACK}{gray}{0}
	\definecolor{WHITE}{gray}{1}
	\definecolor{RED}{rgb}{1,0,0}
	\definecolor{GREEN}{rgb}{0,1,0}
	\definecolor{BLUE}{rgb}{0,0,1}
	\definecolor{CYAN}{cmyk}{1,0,0,0}
	\definecolor{MAGENTA}{cmyk}{0,1,0,0}
	\definecolor{YELLOW}{cmyk}{0,0,1,0}
}

\@ifundefined{definecolor}
{\usepackage{color}}{}
\@ifundefined{definecolor}
{\usepackage{color}}{}
\makeatother

\makeatother

\usepackage{babel}
\begin{document}
	
\title{Next-to-leading order corrections to the spin-dependent energy spectrum of hadrons from polarized top quark decay in the general two Higgs doublet model}

\author{S. Mohammad Moosavi Nejad$^{a,b}$}
\email{mmoosavi@yazd.ac.ir}
	
\author{S. Abbaspour$^a$}

	\affiliation{$^{(a)}$Faculty of Physics, Yazd University, P.O. Box
		89195-741, Yazd, Iran\\	
       $^{(b)}$School of Particles and Accelerators,
		Institute for Research in Fundamental Sciences (IPM), P.O.Box
		19395-5531, Tehran, Iran}

	\date{\today}
	
\begin{abstract}
	
In recent years, searches for the light and heavy charged Higgs bosons have been  done by the ATLAS and the CMS collaborations at the Large Hadron Collider (LHC) in proton-proton collision. Nevertheless, a definitive search is a program that still has to be carried out at the LHC. The experimental observation of charged Higgs bosons would indicate physics beyond the Standard Model.
In the present work, we study the scaled-energy distribution of bottom-flavored mesons ($B$) inclusively produced in polarized top quark decays into a light charged Higgs boson  and a massless bottom quark at next-to-leading order in  the two-Higgs-doublet model; $t(\uparrow)\to bH^+\to BH^++X$.
This spin-dependent energy distribution is  studied in a specific  helicity coordinate system where the polarization vector of the top quark is measured with respect to the direction of the Higgs momentum. The study of these energy distributions could be considered as a new channel to search for the charged Higgs bosons at the LHC. For our numerical analysis and phenomenological predictions, we restrict ourselves to the unexcluded regions of the MSSM $m_{H^+}-\tan\beta$ parameter space  determined by the recent results of  the CMS \cite{CMS:2014cdp} and ATLAS \cite{TheATLAScollaboration:2013wia} collaborations.

\end{abstract}


\maketitle

\section{Introduction}
\label{sec:intro}

The electroweak symmetry breaking in the standard model (SM) of particle physics is described with the Higgs mechanism.
In 2012, the SM Higgs boson with a mass of approximately $125$~GeV was discovered by the CMS and ATLAS experiments \cite{Aad:2012tfa,Chatrchyan:2012xdj} at the CERN  Large Hadron Collider (LHC).  Although the current LHC Higgs data are consistent with the SM, there is still the possibility that the observed Higgs state could be part of a model with an extended Higgs sector.
Models including an extended Higgs sector are constrained by the measured mass, charge-parity (CP) quantum numbers, and production rates of the new boson. The discovery of another heavy scalar boson, neutral or charged, would clearly represent unambiguous evidence for the presence of new physics beyond the standard model. 

Charged Higgs bosons are predicted in models with at least two Higgs doublets.   The simplest of such models is known as the two-Higgs-doublet model (2HDM) \cite{Lee:1973iz} where the Higgs sector of the SM is extended, typically by adding an extra  doublet of complex Higgs fields. After spontaneous symmetry breaking, the particle spectrum of this model includes five physical Higgs bosons: light and heavy CP-even Higgs bosons h and H with $m_h<m_H$, a CP-odd Higgs boson A, plus two charged Higgs bosons $H^\pm$ \cite{Djouadi:2005gj}.
The production mechanisms and decay modes of  charged Higgs bosons depend on their masses, $m_{H^\pm}$. At hadron colliders, a charged Higgs boson can be produced through several channels. For light charged Higgs bosons that their masses are smaller than the difference between the mass of top ($m_t$) and the bottom quark ($m_b$), $m_{H^\pm}< m_t-m_b$, the primary production mechanism is through the decay of a top quark $t\to bH^+$ \cite{Gunion}. Then, in this case, the light
charged Higgs bosons are produced most frequently via $t\bar{t}$ production.
At the LHC, one expects a cross section $\sigma(pp\to t\bar{t}X)\approx 1$ (nb) at design energy $\sqrt{S}=14$ TeV  \cite{Langenfeld:2009tc}. With the LHC design luminosity of $10^{34} cm^{-2}s^{-1}$ in each of the four experiments, it is expected to produce a $t\bar t$ pair per second.  Thus, the LHC is  a superlative top factory which  allows  one to search for the charged Higgs boson  in the subsequent decay products of the top pairs $t\bar t\rightarrow H^\pm H^\mp b\bar b$ and  $t\bar t\rightarrow H^\pm W^\mp b\bar b$ 
when $H^\pm$ decays into $\tau$ lepton and neutrino.
See also Ref.~\cite{Aoki:2011wd} for a review of all available  production modes of light charged Higgs bosons at the LHC in 2HDMs.

The Large Electron-Positron (LEP) collider experiments have determined a model independent low limit  of 78.6 GeV on the charged Higgs mass \cite{Heister:2002ev,Abdallah:2003wd,Achard:2003gt,Abbiendi:2008aa} at a $95\%$ confidence level. The most sensitive $95\%$ confidence level upper limits on the branching fraction $B(t\to bH^+)$ have been determined by the ATLAS and CMS experiments for the mass range $m_{H^+}=80-160$~GeV.  More details can also be found in \cite{Khachatryan:2015qxa}.
We shall discuss about the recent results on a search for a charged Higgs boson by the CMS \cite{CMS:2014cdp} and ATLAS \cite{TheATLAScollaboration:2013wia} collaborations in Sec.~\ref{sec:two}.

The primary purpose of this paper is the evaluation of the next-to-leading order (NLO) QCD corrections to the differential partial decay width ($d\Gamma/dx_i$) of a top quark into a charged Higgs boson and a bottom quark, $t\to bH^+$, where $x_i$ stands for the scaled-energy fraction of the b-quark or the gluon emitted at NLO (see Eq.~(\ref{scale})).
These differential decay widths, which are presented for the first time, are needed to obtain the energy spectrum of B-mesons through top decays.
More detail will be discussed in Sec.~\ref{sec:two}.\\
The $\alpha_s$-order corrections to the top quark decay width, $\Gamma(t \to bH^+)$, were previously computed in \cite{kadeer}  for the polarized top quark and in \cite{Ali:2009sm,Czarnecki,Liud,Li:1990cp} for the unpolarized one. 
In \cite{MoosaviNejad:2011yp}, we calculated the unpolarized  differential decay width $d\Gamma(t\to bH^+)/dx_b$ and showed that our  result after integration over $x_b$ ($0\le x_b\le 1$) is in complete agreement with Refs.~\cite{Ali:2009sm,Czarnecki,Liud} and the corrected version of \cite{Li:1990cp}.
In the present work, to ensure our calculations we check that our  result for the polarized differential decay width $d\Gamma(t(\uparrow)\to bH^+)/dx_b$   is in complete agreement with the result presented in \cite{kadeer} for the polarized decay width $\Gamma(t(\uparrow)\to bH^+)$,  if one integrates over $x_b$, i.e. $\Gamma=\int_0^1 dx_b d\Gamma(t(\uparrow)\to bH^+)/dx_b$. 

On the other hand, the b-quark produced from the top quark decay hadronizes  before it decays, therefore each b-jet contains a bottom-flavored hadron which most of the times is a B-meson, $b\to B+X$. Therefore, the decay process $t\to bH^+(+g)\to BH^++X$ is of prime importance and it is an urgent task to predict its partial decay width ($d\Gamma/dx_B$) as reliably as possible. In fact, one of the proposed ways to search for the charged Higgs bosons at the LHC is the study of the energy distribution of B-mesons inclusively produced in the polarized/unpolarized top quark decays. 
In Ref.~\cite{MoosaviNejad:2011yp}, we studied  the energy spectrum of the bottom-flavored mesons in unpolarized top quark decays into a charged-Higgs boson and a massless bottom quark at NLO in the 2HDM. In the present work we study the energy distribution of B-meson produced through the polarized top  decay $t(\uparrow)\to BH^++X$ at NLO, and compare it to the unpolarized one. For our numerical analysis and our phenomenological predictions, we restrict ourselves to the unexcluded regions of the  $m_{H^+}-\tan\beta$ parameter space  determined by the recent results of  the CMS \cite{CMS:2014cdp} and ATLAS \cite{TheATLAScollaboration:2013wia} collaborations.\\
The top quark polarization can be studied by the angular correlations between the top spin and its decay product momenta so that these spin-momentum correlations will enable  us to detailed study of the top decay mechanism.

Since, highly polarized top quarks will become available at hadron colliders through single top production, at the $33\%$ level of the top pair production rate \cite{Mahlon:1996pn,Espriu:2002wx}, and also at future $e^+e^-$ colliders these measurements of the decay rates will be important to future tests of the Higgs coupling in the minimal supersymmetric SM (MSSM). 

This paper is organized as follows.
In Sec.~\ref{sec:one}, we present our analytical results of the ${\cal O}(\alpha_s)$ QCD corrections to the tree-level rate of $t(\uparrow)\rightarrow bH^+$. We work in the massless scheme where the  b-quark mass is neglected from the beginning  but the arbitrary value of $m_{H^+}$  is retained.
In Sec.~\ref{sec:two}, we present our numerical analysis of inclusive production of a meson from polarized top quark decay considering the factorization theorem and DGLAP equations. We shall compare our result with the one from the unpolarized top decay.
In Sec.~\ref{sec:three},  our conclusions are summarized.

\section{Parton level results in the general two Higgs doublet model}
\label{sec:one}

In this section, assuming the condition $m_t>m_b+m_{H^+}$ we study the NLO radiative corrections to the partial decay width 
\begin{eqnarray}\label{born}
t(\uparrow)\rightarrow b+H^+,
\end{eqnarray}
in the general 2HDM, where  $H_1$ and $H_2$ are the doublets  whose vacuum expectation values respectively give masses to the down and up type quarks. If we label the vacuum expectation values of the fields $H_1$ and $H_2$ as $\textbf{v}_1$  and  $\textbf{v}_2$, respectively,  one has $\textbf{v}_1^2+\textbf{v}_2^2=(\sqrt{2} G_F)^{-1}$ where $G_F$ is the Fermi's constant. The ratio of the two  values $\textbf{v}_1$  and  $\textbf{v}_2$ is a free parameter and one can define the angle $\beta$ to parametrize it, i.e. $\tan\beta=\textbf{v}_2/\textbf{v}_1$. 
Also, a linear combination of the charged components of $H_1$ and $H_2$ gives the observable charged Higgs $H^\pm$, i.e. $H^\pm=H_2^\pm\cos\beta-H_1^\pm\sin\beta$.\\
In a general two Higgs doublet model, in order to avoid a
tree-level flavor-changing neutral currents, the generic
Higgs coupling to all quarks should be restricted.  In fact, one should not couple the same Higgs doublet to up- and down-type quarks simultaneously.
Therefore, we limit ourselves to specific models which  naturally stop these problems by restricting the Higgs coupling.
There are, then, two possibilities (two models)  for the two Higgs doublets to couple to the fermions.\\
In \textbf{model 1}  (first possibility), one of the Higgs doublets ($H_1$) couples to all bosons and the remaining doublet $H_2$ couples to all the quarks.  In this model, the Yukawa couplings between  the top quark, the bottom quark and the charged Higgs boson are given by \cite{GHK}
\begin{eqnarray}\label{modelfirst}
L_1&=&\frac{g_{_W}}{2\sqrt{2}m_W}V_{tb}\cot\beta \bigg\{H^+\bar{t}\big[m_t(1-\gamma_5)-
\nonumber\\
&&m_b(1+\gamma_5)\big]b\bigg\}+H.c,
\end{eqnarray}
where the weak coupling factor $g_W$
is  related to the  Fermi coupling constant   by $g_W^2=4\sqrt{2} m_W^2 G_F$ and $V_{tb}\approx 1$ is the 33 entry of the CKM matrix.

In \textbf{model 2}  (second possibility),  the doublet $H_1$ couples to
the right-handed down-type quarks and the  $H_2$ couples to the right-handed up-type quarks  ($u_R, c_R, t_R$). In  model 2, the interaction Lagrangian leads to an $H^+b\bar{t}$  vertex as
\begin{eqnarray}\label{modelsecond}
L_{2}&=&\frac{g_{_W}}{2\sqrt{2}m_W}V_{tb} \bigg\{H^+\bar{t}\big[m_t\cot\beta(1-\gamma_5)+
\nonumber\\
&&m_b\tan\beta(1+\gamma_5)\big] b\bigg\}+H.c .
\end{eqnarray}
These models are often known as \textbf{Type-I} and \textbf{Type-II} 2HDM scenarios.
The MSSM \cite{Fayet:1974pd,Fayet:1976et,Fayet:1977yc,Dimopoulos:1981zb} is a special case of a \textbf{Type-II} 2HDM scenario.\\
Note that, in type-II 2HDM there is a charged Higgs mass lower limit of  $m_H\approx 480$~GeV at $95\%$  confidence level (CL)\cite{Misiak:2015xwa}. However, this limit is not imposed on a supersymmetric version of type-II (e.g. MSSM). Therefore in this paper, we work on type-I or MSSM as a type-II model.

Generally, the dynamics of the current-induced $t\rightarrow b$ transition is embodied in the following hadron tensor
\begin{eqnarray}\label{tensor}
H^{\mu\nu}&&\propto \sum_{X_b}\int d\Pi_f\delta^4(p_t-p_{_{H^+}}-p_{_{X_b}})\nonumber\\
&&\times <t(p_t,s_t)|J^{\nu\dag}|X_b><X_b|J^{\mu}|t(p_t,s_t)>,\nonumber\\
\end{eqnarray}
where $d\Pi_f$ refers to the Lorentz-invariant phase space factor and $s_t$ stands for the top quark spin.  At NLO approximation, we consider only two types of intermediate states in Eq.~(\ref{tensor}), i.e., $|X_b>=|b>$ for the Born level
term and ${\cal O}(\alpha_s)$ one-loop contributions and $|X_b>=|b+g>$ for the ${\cal O}(\alpha_s)$ tree graph contribution.  
In the SM, the weak current is given by $J^\mu\propto \bar{\psi}_b\gamma^\mu(1-\gamma_5)\bar{\psi}_t $ while 
in the 2HDM,  considering the interaction Lagrangians (\ref{modelfirst}) and (\ref{modelsecond})  the current is expressed as $J^{\mu}\propto \bar{\psi}_b(a+b\gamma_5)\bar{\psi}_t$ in which the coupling factors are
\begin{eqnarray}\label{model1}
\textbf{model 1}:\quad a&=&\frac{g_{_W}}{2\sqrt{2}m_W}V_{tb}(m_t-m_b)\cot\beta,\nonumber\\
b&=&\frac{g_{_W}}{2\sqrt{2}m_W}V_{tb}(m_t+m_b)\cot\beta,
\end{eqnarray}
or
\begin{eqnarray}\label{model2}
\textbf{model 2}:\quad
a&=&\frac{g_{_W}}{2\sqrt{2}m_W}V_{tb}(m_t \cot\beta+m_b\tan\beta),\nonumber\\
b&=&\frac{g_{_W}}{2\sqrt{2}m_W}V_{tb}(m_t \cot\beta-m_b\tan\beta).\nonumber\\
\end{eqnarray}
The decay process (\ref{born}) is analyzed in the rest frame of the top quark where the three-momentum $\vec{P}_H$ of the $H^+$ boson points into the direction of the positive $z$-axis and the polar angle $\theta_P$ is defined as the angle between the polarization vector $\vec{P}_t$ of the top quark and the $z$-axis (see Fig.~\ref{lo}). Here, we follow the notation of Ref.~\cite{MoosaviNejad:2011yp} where we discussed the NLO radiative corrections to the partial decay rate of unpolarized top quarks.\\
The angular distribution of the  differential decay width $d\Gamma/dx$ of a polarized top quark is given by the following simple expression
to clarify the correlation between the polarization of the top quark and its decay products
\begin{eqnarray}\label{widthdefine}
\frac{d^2\Gamma}{dx_b d\cos\theta_P}=\frac{1}{2}(\frac{d\Gamma^{unpol}}{dx_b}+P\frac{d\Gamma^{pol}}{dx_b}\cos\theta_P),
\end{eqnarray}
where $P$ is the degree of the top quark polarization with $0\leq P\leq 1$ such that  $P=1$ corresponds to $100\%$ top quark polarization and
$P=0$ corresponds to an unpolarized top quark.
In this equation, following Refs.~\cite{MoosaviNejad:2011yp,Kniehl:2012mn} we have defined the   scaled-energy fraction of the b-quark as
\begin{eqnarray}\label{scale}
x_b=\frac{E_b}{E_b^{max}}=\frac{2E_b}{m_t(1+R-y)},
\end{eqnarray}
where the dimensionless parameters $y=m_H^2/m_t^2$ and $R=m_b^2/m_t^2$ are defined. By neglecting the b-quark mass one has $x_b=2E_b/(m_t(1-y))$ so that $0\leq x_b \leq 1$. 
In Eq.~(\ref{widthdefine}),  $d\Gamma^{pol}/dx_b$ stands for the  polarized differential rate and $d\Gamma^{unpol}/dx_b$ refers to the unpolarized one which is extensively calculated in \cite{MoosaviNejad:2011yp} up to NLO.\\
In the following,  we express the technical detail of our calculation for the ${\cal O}(\alpha_s)$ radiative corrections to the tree-level decay rate of $t(\uparrow)\rightarrow b+H^+$ using dimensional regularization.

\subsection{Born level rate of $t(\uparrow)\rightarrow bH^+$}
\begin{figure}
	\begin{center}
		\includegraphics[width=0.8\linewidth,bb=20 10 400 190]{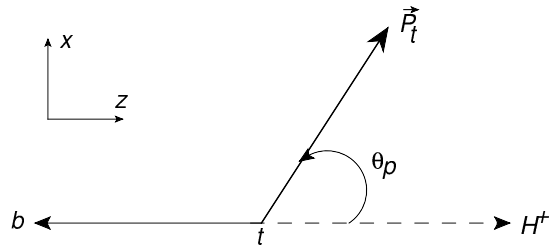}
		\caption{\label{lo}%
			Definition of the polar angle $\theta_P$ in the top quark rest frame. $\vec{P}_t$ is the polarization vector of the top quark.}
	\end{center}
\end{figure}
\begin{figure}
	\begin{center}
		\includegraphics[width=0.55\linewidth,bb=40 10 180 170]{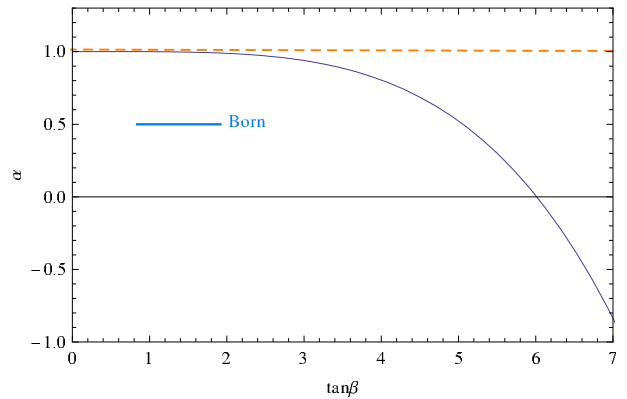}
		\caption{\label{ratio}%
			Ratio of polarized decay rates at the Born-level for two models ($\alpha=\Gamma_0^{2,pol}/\Gamma_0^{1,pol}$) as a function of $\tan\beta$.}
	\end{center}
\end{figure}
It is straightforward to compute the Born term contribution to the partial decay rate of the polarized top quark in the 2HDM. According to  the interaction Lagrangians (\ref{modelfirst}) and (\ref{modelsecond}), the coupling of the charged-Higgs boson to the bottom and top quarks  can either be expressed as a superposition of  scalar and pseudoscaler coupling factors or as a superposition of right- and left-chiral coupling factors \cite{GHK}. Therefore, the Born term amplitude of the process (\ref{born}) is given by 
\begin{eqnarray}\label{b1}
M_0=\bar{u_b}(a\boldsymbol{1}+b\gamma_5)u_t=\bar{u_b}\{g_t\frac{1+\gamma_5}{2}+g_b\frac{1-\gamma_5}{2}\}u_t,
\end{eqnarray}
where, $a$ and $b$ depend on the model and given in (\ref{model1}) and (\ref{model2}). One also has  $g_t=a+b$ and $g_b=a-b$.\\
 For the amplitude squared, one has
\begin{eqnarray}\label{toftof}
|M_0|^2&=&\sum_{s_b}M_0^\dag M_0=2(p_b\cdot p_t)(a^2+b^2)+\nonumber\\
&&2(a^2-b^2)m_bm_t+4ab m_t(p_b\cdot s_t),
\end{eqnarray}
where we replaced $\sum_{s_t}u(p_t, s_t)\bar{u}(p_t, s_t)=(\displaystyle{\not}{p}_t+m_t)$ in the unpolarized Dirac string by $u(p_t, s_t)\bar{u}(p_t, s_t)=(1-\gamma_5 \displaystyle{\not}{s}_t )(\displaystyle{\not}{p}_t+m_t)/2$ in the polarized state.\\
Considering Fig.~\ref{lo}, the polarization four-vector of the top quark in the top rest frame reads; $s_t=P(0;\sin\theta_P \cos\phi_P, \sin\theta_P\sin\phi_P,\cos\theta_P)$ so that  one has $p_b\cdot s_t=P(|\vec{p}_b|\cos\theta_P)$.
 Therefore, the polarized tree-level decay width reads
\begin{eqnarray}\label{gammatree}
\Gamma_0^{pol}&=&\frac{m_t}{8\pi}\lambda(1,\frac{m_b^2}{m_t^2},
\frac{m_{H^+}^2}{m_t^2})(ab),
\end{eqnarray}
where $\lambda(x,y,z)=(x-y-z)^2-4y z$ is the  K\"all\'en function.
The above result is in complete agreement with  Refs.~\cite{kadeer,Liud}.
The unpolarized Born-level rate can be found in our previous work \cite{MoosaviNejad:2011yp}.
In (\ref{gammatree}) for the product of two coupling factors, in the model $1$, one has
\begin{eqnarray}\label{haselmodel1}
ab=\frac{G_F}{\sqrt{2}} |V_{tb}|^2 (m_t^2-m_b^2)\cot^2\beta,
 \end{eqnarray}
and for the model 2,
\begin{eqnarray}\label{haselmodel2}
ab=\frac{G_F}{\sqrt{2}} |V_{tb}|^2 (m_t^2 \cot^2\beta-m_b^2 \tan^2\beta).
 \end{eqnarray}
Considering (\ref{gammatree}) and (\ref{haselmodel2}), it is seen that  in the model 2 the rate becomes zero when   $\tan\beta=\sqrt{m_t/m_b}\approx 6$ if we take $m_t=172.9$~GeV and $m_b=4.78$~GeV.\\
Defining $\alpha=\Gamma_0^{2,pol}/\Gamma_0^{1,pol}$ as a ratio of polarized Born widths in the models 1 and 2, in Fig.~\ref{ratio} we plot this ratio as a function of $\tan\beta$. Note that, this ratio is independent of the charged Higgs boson mass and for $\tan\beta<4$ (with $\alpha\approx 1$) the Born rates are the same in both models. 
$\alpha$ is positive/negative for small/large values of $\tan\beta$ and goes through zero for $\tan\beta=6.01$.
\\
In this work, we adopt  the massless scheme or  Zero-Mass Variable-Flavor-Number (ZM-VFN) scheme \cite{jm} where the zero mass parton approximation is also applied  to the bottom quark. 
In \cite{kadeer}, it is shown that the $m_b=0$ approximation can be quite good in both models, see Figs.~5a and b of this reference.

In the limit of vanishing b-quark mass, the tree-level decay width is simplified to
\begin{eqnarray}\label{gammaaa}
\Gamma_0=\frac{m_t}{8\pi}(1-\frac{m_{H^+}^2}{m_t^2})^2(ab).
\end{eqnarray}
In the following,  in a detailed discussion we calculate the ${\cal O}(\alpha_s)$ QCD corrections to the Born-level decay rate of $t\rightarrow bH^+$ and
we present, for the first time, the analytical parton-level expressions for $d\Gamma(t(\uparrow)\rightarrow BH^++X)/dx_B$ at
NLO in the ZM-VFN scheme.

\subsection{${\cal O}(\alpha_s)$ virtual corrections}\label{virtual}

The ${\cal O}(\alpha_s)$ one-loop vertex corrections to the $tbH^+$-vertex
arise from the emission and absorption of the virtual gluons from top and bottom quark legs in Feynman diagrams. Considering the interaction of the quark fields $q(x^\mu)$ with gluons which  includes a vector-like coupling as
\begin{eqnarray}\label{lag2}
g_s\bar{q}_i(x)\gamma^\mu T^a_{ij} q_j(x) G^a_\mu (x),
\end{eqnarray}
one can extract the Feynman rules to calculate the virtual radiative corrections. In (\ref{lag2}), $g_s$ is the strong coupling constant, $a=1,2,\cdots, 8$ is the QCD color index of gluons so  for the SU(3) generator $T^a$ one has $Tr(T^aT^a)=4$.\\
In the massless-scheme where $m_b=0$ is considered, the virtual one-loop corrections consist of both infrared (IR) and ultraviolet (UV) divergences  in which, for example, the  UV-divergences appear  when the integration region of the internal momentum of the virtual gluon goes to infinity.  Here, we adopt the "on-shell" mass renormalization
scheme and use dimensional regularization to regulate all singularities.  In this scheme, all divergences  are regularized in $D =4-2\epsilon$ (with $\epsilon\ll 1$) space-time dimensions to become single poles in $\epsilon$.

Considering the scaled-energy variable (\ref{scale}), which is now simplified in the massless scheme as 
\begin{eqnarray}\label{scalenew}
x_b=\frac{2E_b}{m_t(1-y)},
\end{eqnarray}
 the contribution of virtual corrections into the doubly differential decay width (\ref{widthdefine}) is given by
\begin{eqnarray}
\frac{d^2\Gamma^{\textbf{vir}}_b}{dx_b d\cos\theta_P}=\frac{|M^{\textbf{vir}}|^2}{32\pi m_t}(1-y)\delta(1-x_b),
\end{eqnarray}
where, $|M^{\textbf{vir}}|^2=\sum_{s_b}(M_0^{\dagger} M_{loop}+M_{loop}^{\dagger} M_0)$. The Born amplitude $M_0$ is given in (\ref{b1}) and following Refs.~\cite{Czarnecki,Liud}, the renormalized amplitude of the virtual corrections  is written as
\begin{eqnarray}\label{tiftif}
M_{loop}=\bar{u_b}(\Lambda_{ct}+\Lambda_l)(a+b\gamma_5)u_t,
\end{eqnarray}
where $\Lambda_l$ stands for the one-loop vertex correction and $\Lambda_{ct}$ refers to the counter term of the vertex. The analytical form of the counter term (including the mass and the wave-function renormalizations of the top and bottom quarks), and the one-loop vertex correction $\Lambda_l$ can be found in \cite{MoosaviNejad:2011yp} when the massless-scheme is applied. For the massive scheme (where $m_b\neq 0$) these forms can be found in \cite{MoosaviNejad:2012ju}. \\
Note that,  after summing all virtual corrections up all UV-divergences  are canceled but the IR-singularities are remaining which, from now on, we label them by $\epsilon$.\\
Therefore, the virtual corrections to the differential decay width (\ref{widthdefine}) is presented by
\begin{eqnarray}\label{virtualfinal}
\frac{d^2\Gamma^{\textbf{vir}}}{dx_b d\cos\theta_P}=\frac{1}{2}\bigg\{\frac{d\Gamma^{\textbf{vir,unpol}}}{dx_b}+P\frac{d\Gamma^{\textbf{vir,pol}}}{dx_b}\cos\theta_P\bigg\},
\end{eqnarray}
where $d\Gamma^{\textbf{vir,unpol}}/dx_b$ is given in \cite{MoosaviNejad:2011yp} and for the polarized rate, normalized to the polarized Born width (\ref{gammaaa}), one has
\begin{eqnarray}\label{saman}
\frac{1}{\Gamma_0}\frac{d\Gamma^{\textbf{vir,pol}}}{dx_b}&=&\frac{\alpha_s(\mu_R)}{2\pi}C_F\delta(1-x_b)\big(-\frac{1}{\epsilon^2}+\frac{F}{\epsilon}-\frac{F^2}{2}+\nonumber\\
&&(\frac{2}{y}-5)\ln(1-y)-2Li_2(y)-\frac{7}{8}-\frac{\pi^2}{12}\big).\nonumber\\
\end{eqnarray}
Here, $C_F=(N_c^2-1)/(2N_c)=4/3$ for $N_c=3$ quark colors, $Li_2(y)$ is the Spence function and the term $F$ reads
\begin{eqnarray}
F=2\ln(1-y)-\ln\frac{4\pi\mu_F^2}{m_t^2}+\gamma_E-\frac{5}{2},
\end{eqnarray}
where  $\mu_F$ is the factorization scale and $\gamma_E=0.5772\cdots$ is the Euler constant.\\
The renormalized virtual one-loop correction (\ref{saman}) is in complete agreement with \cite{kadeer}. Although, this comparison is not so straightforward, because in \cite{kadeer} authors regularized the UV singularities using the D-dimensional regularization scheme (as we have done) but to regulate the IR divergences they introduced a finite (small) gluon mass $m_g\neq 0$ in the gluon propagator. Then, to compare the extracted results one has to consider  the replacement; $1/\epsilon-\gamma_E+\ln(4\pi \mu_F^2/m_t^2)\to \ln (m_g^2/m_t^2)$.
However, all the logarithmic gluon mass dependence or/and the singular terms in the form of $1/\epsilon$ resulting from the different regularization procedures must be canceled out when the virtual and tree-graph contributions are summed up.

\subsection{Tree-graph contributions}
\label{real}
\begin{figure}
	\begin{center}
		\includegraphics[width=0.9\linewidth,bb=20 10 450 290]{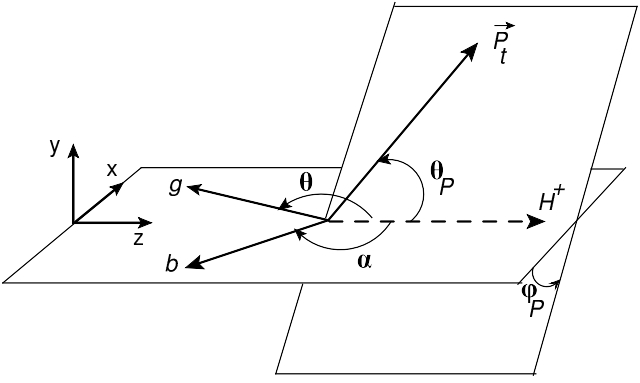}
		\caption{\label{nlo}%
			Definition of the polar angle $\theta_P$ in the helicity coordinate system selected for the NLO QCD corrections. $\vec{P}_t$ is the polarization vector of the top quark.}
	\end{center}
\end{figure}
In the rest frame of a top quark decaying into a bottom quark, a Higgs boson and a gluon, the outgoing particles define an event plane. Relative to this plane we can define the spin direction of the polarized top quark. Therefore, for the NLO analysis of the spin-momentum correlation between the top quark polarization vector and the momenta of its decay products we apply the helicity coordinate system shown in Fig.~\ref{nlo}. In this system the polarization vector of the top quark is evaluated relative to the Higgs boson 3-momentum which points to the direction of the positive $z$-axis.\\
The QCD NLO contribution to the differential decay rate results from the square of the amplitudes as
$|M_0|^2$ (\ref{toftof}), $|M^{\textbf{vir}}|^2$ (\ref{tiftif}) and $|M^{\textbf{real}}|^2=M^{\textbf{real} \dagger}\cdot M^{\textbf{real}}$, where $M^{\textbf{real}}$ stands for the real gluon (tree-graph) contribution, $t(\uparrow)\to bH^++g$, which reads
\begin{eqnarray}\label{finfin}
M^{\textbf{real}}&=&g_s\frac{\lambda^a}{2}\bar u(p_b, s_b)\big\{\frac{2p_t^\mu-
\displaystyle{\not}p_g \gamma^\mu}{2p_t \cdot p_g}
\\
&&-\frac{2p_b^\mu+\gamma^\mu \displaystyle{\not}p_g}
{2p_b\cdot p_g}\big\}(a\textbf{1}+b\gamma_5) u(p_t, s_t)\epsilon_{\mu}^{\star}(p_g,r),\nonumber
\end{eqnarray}
where $\epsilon(p_g,r)$ stands for the polarization vector of the  emitted real gluon with the momentum $p_g$ and spin $r$. In (\ref{finfin}), the first and second terms refer to real gluon emission from the top quark and the bottom quark, respectively.
As before, in order to regulate the IR-divergences, which arise from the soft- and collinear-gluon emissions, we work in  $D-$dimensions.
In this scheme, the differential decay rate for the real emission contribution is given by
\begin{eqnarray}\label{moozmooz}
d\Gamma^{\textbf{real}}=\frac{\mu_F^{2(4-D)}}{2m_t}|M^{\textbf{real}}|^2dR_3(p_t, p_b, p_g, p_{_{H^+}}),
\end{eqnarray}
where, the phase space element $dR_3$ is
\begin{eqnarray}\label{ahah}
\frac{d^{D-1}\bold{p}_b}{2E_b}\frac{d^{D-1}\bold{p}_H}{2E_H}\frac{d^{D-1}\bold{p}_g}{2E_g}
(2\pi)^{3-2D}\delta^D(p_t-\sum_{g,b,H} p_f).\nonumber\\
\end{eqnarray}
To evaluate the real doubly differential decay rate   
normalized to the polarized Born width (\ref{gammaaa}), i.e.  $1/\Gamma_0\times d^2\Gamma^{\textbf{real}}/(dx_bd\cos\theta_P)$, 
we fix the momentum of b-quark in (\ref{moozmooz}) and integrate over
the energy of the $H^+$-boson which ranges as
\begin{eqnarray}
m_t\frac{y+[1-x_b(1-y)]^2}{2[1-x_b(1-y)]}\leq E_H \leq  m_t\frac{1+y}{2}.
\end{eqnarray}
To compute the  angular distribution of differential width, 
the angular integral in (\ref{ahah}) has to be written as $d^{D-1}\bold{p}_{_H}=\bold{p}_{_H}^{D-2}d\bold{p}_{_H}d\Omega_H$ in which
\begin{eqnarray}
d\Omega_H=-\frac{2\pi^{\frac{D}{2}-1}}{\Gamma(\frac{D}{2}-1)}(\sin\theta_{P})^{D-4} d\cos\theta_{P}.
\end{eqnarray}
Therefore, the polarized doubly differential width reads
\begin{eqnarray}\label{mohsen}
\frac{d^2\Gamma^{\textbf{real,pol}}}{dx_b d\cos\theta_{P}}&=&A x_b^{D-4}|M^{real}|^2(1-\cos^2\alpha)^{\frac{D-4}{2}}\times\nonumber\\
&&\delta(\cos\alpha-b)dE_H d\cos\alpha,
\end{eqnarray}
where the angles $\theta_P$ and $\alpha$ are defined in Fig.~\ref{nlo}, and 
\begin{eqnarray}\label{panj}
A=\mu_F^{2(4-D)}(p_H m_t)^{D-4}\frac{(1-y)^{D-3}}{2^{3D-4}\pi^{D-1}\Gamma^2(\frac{D}{2}-1)}.
\end{eqnarray}
In the equation above, $b=(m_t^2+m_H^2-2m_t(E_b+E_H)+2E_bE_H)/(2E_bp_H)$ and 
$p_H=\sqrt{E_H^2-m_H^2}$ is the 3-momentum of the Higgs boson.\\
Considering Fig.~\ref{nlo}, the relevant scalar products evaluated in the top rest frame are
\begin{eqnarray}
p_H\cdot s_t&=&-P(p_H\cos\theta_{P}),\nonumber\\
p_b\cdot s_t&=&-P(E_b\cos\alpha\cos\theta_{P}),\\
p_b\cdot p_H&=&E_b(E_H-p_H\cos\alpha),\nonumber
\end{eqnarray}
and $p_t\cdot s_t=0$. Here $P$ refers to the polarization degree of the top quark.

It should be noted that, since the real correction contribution includes the pole $\propto 1/\epsilon$, therefore, to get the correct finite terms in the  normalized differential distributions the Born width $\Gamma_0$ (\ref{gammaaa}) must be evaluated 
in the dimensional regularization at ${\cal O}(\epsilon)$, i.e.
$\Gamma_0\rightarrow \Gamma_0\{1-\epsilon
\big[2\ln(1-y)-2\ln 2+\gamma_E-\ln(4\pi\mu_F^2/m_t^2)\big]\}$.

As a last technical point; when one integrates over the phase 
space for the real gluon radiation, terms of the form $(1-x_b)^{-1-2\epsilon}$ arise which are due to the radiation of a soft gluon in top decay. In fact, the limit of $E_g\to 0$ corresponds to the limit $x_b\to 1$. Therefore, we use the following expression \cite{Corcella:1}
\begin{eqnarray}
(1-x_b)^{-1-2\epsilon}&&=-\frac{1}{2\epsilon}\delta(1-x_b)+\bigg(\frac{1}{1-x_b}\bigg)_+\nonumber\\
&&-2\epsilon \bigg(\frac{\ln(1-x_b)}{1-x_b}\bigg)_+,
\end{eqnarray}
where the plus distribution is defined as 
\begin{eqnarray}
\int_0^1 (f(x_b))_{_+}h(x_b)dx_b=\int_0^1f(x_b)[h(x_b)-h(1)]dx_b.
\end{eqnarray}


\subsection{Analytical results for differential decay rates $d\Gamma/dx_i$ at parton level}

According to Eq.~(\ref{widthdefine}), the ${\cal O}(\alpha_s)$ correction to the angular distribution of partial decay rates 
is obtained by  summing the Born, the virtual and the real gluon contributions and is given by
\begin{eqnarray}\label{firsttt}
\frac{d^2\Gamma_{\textbf{nlo}}}{dx_b d\cos\theta_P}=\frac{1}{2}\bigg\{\frac{d\Gamma^{\textbf{unpol}}_{\textbf{nlo}}}{dx_b}+P\frac{d\Gamma^{\textbf{pol}}_{\textbf{nlo}}}{dx_b}\cos\theta_P\bigg\}.
\end{eqnarray}
The unpolarized rate $d\Gamma^{\textbf{unpol}}/dx_b$ is given in \cite{MoosaviNejad:2011yp} and for the polarized one, normalized to the Born rate (\ref{gammaaa}), one has
\begin{eqnarray}\label{pol1}
\frac{1}{\Gamma_0}\frac{d\Gamma^{\textbf{pol}}_{\textbf{nlo}}}{dx_b}&=&\delta(1-x_b)+
\frac{C_F\alpha_s}{2\pi}\Big\{[-\frac{1}{\epsilon}+\gamma_E-\ln 4\pi]\nonumber\\
&&\times[\frac{3}{2}\delta(1-x_b)+\frac{1+x_b^2}{(1-x_b)_+}]+T_1\Big\},
\end{eqnarray}
where, by defining $S=(1-y)/2$ (with $y=m_H^2/m_t^2$) one has
\begin{eqnarray}\label{pol11}
T_1&=&\delta(1-x_b)\bigg\{-\frac{3}{2}\ln\frac{\mu_F^2}{m_t^2}+\frac{4S}{y}\ln(1-y)-\frac{7\pi^2}{3}\nonumber\\
&&+2Li_2(1-y)-2Li_2(y)-\frac{y}{S}\ln(y)-4\bigg\}\nonumber\\
&&+2(1+x_b^2)\bigg(\frac{\ln(1-x_b)}{1-x_b}\bigg)_+\nonumber\\
&&+2\frac{1+x_b^2}{(1-x_b)_+}\ln\frac{m_tx_b(1-y)}{\mu_F}+1-x_b-R_2\frac{1+x_b^2}{1-x_b}\nonumber\\
&&-\frac{y}{S(1-x_b)}+\frac{1-S}{S(Sx_b^2-2x_b+2)}+\nonumber\\
&&\frac{|2Sx_b^2-2x_b+1|}{Sx_b^2-2x_b+2}\bigg(\frac{1}{1-x_b}+\frac{1-Sx_b}{S}\bigg)+\nonumber\\
&&R_1\sqrt\frac{Sx_b^2-2x_b+2}{S}\bigg(\frac{1+x_b}{1-x_b}+\frac{x_bS^2-Sx_b-1}{S(Sx_b^2-2x_b+2)}+\nonumber\\
&&\frac{1-S-x_b(1-S)^2}{S(Sx_b^2-2x_b+2)^2}\bigg).\nonumber\\
\end{eqnarray}
Here, we also defined
\begin{eqnarray}
R_1&=&\ln\bigg(1-2S^2x_b^3+4Sx_b^2-3Sx_b-x_b+\nonumber\\
&&\sqrt{S(Sx_b^2-2x_b+2)}|2Sx_b^2-2x_b+1|\bigg),\\
R_2&=&\ln\bigg((1-S)x_b^2-x_b+\frac{1}{2}+\frac{|2Sx_b^2-2x_b+1|}{2}\bigg).\nonumber
\end{eqnarray}
This differential decay rate (\ref{pol1})  after integration over $x_b$ ($0\leq x_b \leq 1$) is in complete agreement with the result presented in \cite{kadeer}.

Since, observable hadrons through top decays can be also produced from the fragmentation of the emitted real gluons, therefore, to obtain 
the most accurate energy spectrum of produced hadrons one has to add the contribution of gluon fragmentation to 
the b-quark one to produce the outgoing hadron.
As shown in \cite{Nejad:2013fba},  the gluon splitting contribution is  
important at a low energy of the observed hadron  so this decreases the size of decay rate at the threshold.
Then, we also need the polarized differential decay rate $d\Gamma^{\textbf{pol}}_{\textbf{nlo}}/dx_g$, where $x_g=2E_g/(m_t (1-y))$ is the scaled-energy fraction of the real gluon, as  in (\ref{scalenew}).
Considering the general form  of the angular distribution (\ref{firsttt}), the unpolarized rate $d\Gamma^{\textbf{unpol}}/dx_g$ is given in \cite{MoosaviNejad:2011yp} and for the polarized one we proceed as follows.
In (\ref{moozmooz}), we fix the momentum of gluon in the three-body phase space and integrate over
the energy of the $H^+$-boson which ranges as
\begin{eqnarray}
m_t\frac{y+[1-x_g(1-y)]^2}{2[1-x_g(1-y)]}\leq E_H \leq  m_t\frac{1+y}{2}.
\end{eqnarray}
Therefore, the polarized doubly differential decay rate is obtained by
\begin{eqnarray}
\frac{d^2\Gamma^{\textbf{pol}}}{dx_g d\cos\theta_{P}}&\propto& x_g^{D-4}|M^{real}|^2(1-\cos^2\theta)^{\frac{D-4}{2}}\times\nonumber\\
&&\delta(\cos\theta-a)dE_H d\cos\theta,
\end{eqnarray}
where, the proportionality coefficient is the same as in (\ref{panj}) and  $\theta$ is the angle between the 3-momentum of the gluon 
and the Higgs boson (see Fig.~\ref{nlo}), whereas $a=(m_t^2+m_H^2-2m_t(E_g+E_H)+2E_gE_H)/(2E_gp_H)$.
The required four-momentum scalar products are 
\begin{eqnarray}
p_H\cdot s_t&=&-P(p_H\cos\theta_{P}),\nonumber\\
p_g\cdot s_t&=&-P(E_g\cos\theta\cos\theta_{P}),\\
p_g\cdot p_H&=&E_g(E_H-p_H\cos\theta).\nonumber
\end{eqnarray}
Therefore the polarized differential width, normalized to the Born rate (\ref{gammaaa}), is expressed as
\begin{eqnarray}\label{pol2}
\frac{1}{\Gamma_0}\frac{d\Gamma^{\textbf{pol}}_{\textbf{nlo}}}{dx_g}&=&
\frac{C_F\alpha_s}{2\pi}\bigg\{\frac{1+(1-x_g)^2}{x_g}(-\frac{1}{\epsilon}+\gamma_E-\ln 4\pi)\nonumber\\
&&+T_2\bigg\},
\end{eqnarray}
where,
\begin{eqnarray}\label{pol22}
T_2&=&\frac{1+(1-x_g)^2}{x_g}\bigg(-B_2+2\ln\frac{2Sx_g(1-x_g)m_t}{\mu_F}\bigg)\nonumber\\
&&+\frac{1-S(1+3x_g)+x_gS^2(2x_g^2+4x_g-3)}{2S^2x_g^2}+\nonumber\\
&&\frac{B_1}{2S^3x_g^2}\bigg(1-S(3x_g+2)-2x_gS^3(x_g^2-2x_g+2)\nonumber\\
&&+2S^2(1+x_g)^2\bigg)+\frac{|2Sx_g^2-2x_g+1|}{2S^2x_g^2(1-2Sx_g)^2}\bigg(6S^3x_g^3\nonumber\\
&&-4S^3x_g^2-x_g(1+8x_g)S^2+5Sx_g+S-1\bigg),
\end{eqnarray}
and
\begin{eqnarray}
B_1&=&\ln\bigg(2S^2x_g^2-2Sx_g-S+1+S|2Sx_g^2-2x_g+1|\bigg)\nonumber\\
&&-\ln(1-2Sx_g),\\
B_2&=&\ln\bigg((1-S)x_g^2-x_g+\frac{1}{2}+\frac{|2Sx_g^2-2x_g+1|}{2}\bigg).\nonumber
\end{eqnarray}

In Eqs.~(\ref{pol1}) and (\ref{pol2}), $T_1$ and $T_2$ are free of all divergences  and to subtract the singularities 
remaining in the polarized differential decay widths, we apply the modified minimal-subtraction $(\overline{MS})$ scheme where, the singularities are absorbed
into the bare fragmentation functions. This renormalizes the fragmentation functions and creates the finite terms of the form $\alpha_s\ln(m_t^2/\mu_F^2)$ in 
the polarized differential widths. Following this scheme, to obtain  the $\overline{MS}$ coefficient functions one  has to subtract from
(\ref{pol1}) and (\ref{pol2}),  the ${\cal O}(\alpha_s)$ term multiplying the  characteristic $\overline{MS}$ constant $(-1/\epsilon+\gamma_E-\ln 4\pi)$\cite{Corcella:1}. \\
In this work we set $\mu_F=m_t$, so that in Eqs.~(\ref{pol11}) and (\ref{pol22}) the terms proportional to $\alpha_s\ln(m_t^2/\mu_F^2)$ vanish.

\section{Numerical results}
\label{sec:two}

Having the parton-level differential decay widths (\ref{pol1}) and (\ref{pol2}), we are now in a situation to present  our phenomenological predictions  
for the scaled-energy ($x_B$) distribution of bottom-flavored hadrons (B) inclusively  produced in polarized top decay in the 2HDM.
To indicate  our predictions for the $x_B$-distribution, we
consider the doubly differential distribution  $d^2\Gamma/(dx_B d\cos\theta_P)$ of the partial width of the decay $t(\uparrow)\rightarrow BH^++X$. Here, as in (\ref{scalenew}),  
$x_B=2E_B/(m_t(1-y))$ is the scaled-energy fraction of the B-hadron in the top quark rest frame, where the B-hadron energy ranges from $E_B^{min}=m_B$ to $E_B^{max}=(m_t^2+m_B^2-m_H^2)/(2m_t)$.

In general case, according to the factorization  theorem of QCD-improved parton model \cite{collins}, the B-hadron energy
distribution  can be expressed as  the convolution of the parton-level spectrum $d\Gamma/dx_a (a=b, g)$ with
the nonperturbative fragmentation function $D_a^B(z, \mu_F)$ which describes the hadronization process $a\to B$ at the scale $\mu_F$, i.e.
\begin{equation}\label{eq:master}
\frac{d\Gamma}{dx_B}=\sum_{a=b, g}\int_{x_a^\text{min}}^{x_a^\text{max}}
\frac{dx_a}{x_a}\,\frac{d\Gamma}{dx_a}(\mu_R, \mu_F) D_a^B(\frac{x_B}{x_a}, \mu_F),
\end{equation}
where,  $\mu_R$ is the renormalization scale related to
the renormalization of the QCD coupling constant and $\mu_F$ is the factorization scale.
We shall use the convention $\mu_R=\mu_F=m_t$ for our results, a choice often made.

\begin{figure}
	\begin{center}
		\includegraphics[width=0.7\linewidth,bb=110 32 490 540]{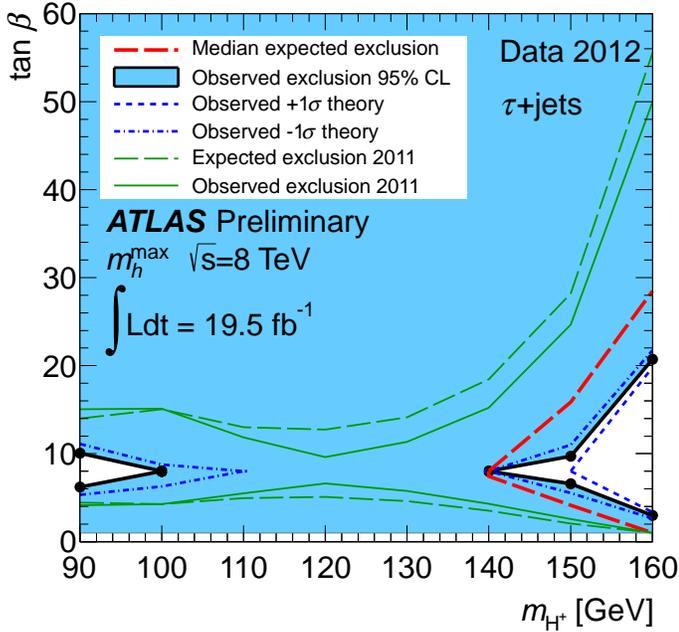}
		\caption{\label{plot}%
			 Exclusion region in the MSSM $\tan\beta-m_{H^+}$ parameter space for $m_{H^+}=80-160$ GeV is shown.
			 The $\pm 1\sigma$ band around the expected limit is also shown. The blue region is excluded. Plot is got from Ref.~\cite{TheATLAScollaboration:2013wia}.}
	\end{center}
\end{figure}

In the MSSM, the mass of $H^\pm$  is strongly correlated with the mass of other Higgs bosons. In this model, the charged Higgs boson mass is restricted at tree-level by $m_{H^+}>m_W$ \cite{Nakamura:2010zzi}, but this restriction does not hold for some regions of parameter space after including radiative corrections.
In \cite{Ali:2009sm}, it is mentioned that 
a $H^\pm$ boson with a mass in the range $80 GeV\leq m_{H^\pm}\leq 160 GeV$ is a logical possibility
and its effects should be searched for in the decays $t\rightarrow  bH^+\rightarrow B\tau^+\nu_\tau+X$.\\
On the other hand, the recent results of a search for evidence of a charged Higgs boson in $19.5-19.7 fb^{-1}$ of proton-proton collision data recorded at $\sqrt{s}=8$~TeV are reported by the CMS \cite{CMS:2014cdp} and the ATLAS \cite{TheATLAScollaboration:2013wia}  experiments at the CERN LHC.
Their results show that the large region in the MSSM $m_{H^+}-\tan\beta$ parameter space for $m_{H^+}=80-160$~GeV is excluded and only some regions of the parameter space are  still unexcluded. These regions along with the $\pm 1\sigma$ band around the expected limit are  shown in Fig.~\ref{plot} which  is taken from Ref.~\cite{TheATLAScollaboration:2013wia}. A same exclusion is reported by the  CMS \cite{CMS:2014cdp} collaboration. However, a definitive search of the charged-Higgses over this part of the $m_{H^+}-\tan\beta$ plane in the MSSM is a program that still has to be carried out and this belongs to the LHC experiments.  

For our numerical analysis, from Ref.~\cite{Nakamura:2010zzi} we adopt the input parameter values
$G_F = 1.16637\times10^{-5}$~GeV$^{-2}$,
$m_t = 172.98$~GeV,
$m_W=80.399$~GeV,
$m_B = 5.279$~GeV, and
$|V_{tb}|=0.999152$.
We evaluate the QCD coupling constant $\alpha_s^{(n_f)}(\mu_R)$ at NLO in the $\overline{\text{MS}}$ scheme using
\begin{eqnarray}\label{alpha}
\alpha^{(n_f)}_s(\mu)=\frac{1}{b_0\log(\mu^2/\Lambda^2)}
\Big\{1-\frac{b_1 \log\big[\log(\mu^2/\Lambda^2)\big]}{b_0^2\log(\mu^2/\Lambda^2)}\Big\},
\nonumber\\
\end{eqnarray}
where $n_f$ is the number of active quark flavors, and $b_0$ and $b_1$ are given by 
\begin{eqnarray}
b_0=\frac{33-2n_f}{12\pi}, \quad  b_1=\frac{153-19n_f}{24\pi^2},
\end{eqnarray}
where $\Lambda$ is the typical QCD scale. Here,  we adopt
$\Lambda_{\overline{\text{MS}}}^{(5)}=231.0$~MeV adjusted such
 that $\alpha_s^{(5)}=0.1184$ for $m_Z=91.1876$~GeV \cite{Nakamura:2010zzi}.
To describe the splitting  $(b, g)\rightarrow B$, we employ the
realistic nonperturbative $B$-hadron  fragmentation functions determined at NLO in the ZM-VFN scheme through a global fit  to
electron-positron annihilation data presented by ALEPH \cite{Heister:2001jg} and OPAL
\cite{Abbiendi:2002vt} at CERN LEP1 and by SLD \cite{Abe:1999ki} at SLAC SLC. Specifically,  for the $b\to B$ splitting a simple power model as; $D_b(z,\mu_F^\text{ini})=Nz^\alpha(1-z)^\beta$
was used at the initial scale $\mu_F^\text{ini}=4.5$~GeV, while the   gluon and light-quark fragmentation functions were generated
via the DGLAP  evolution equations \cite{dglap}.
The fit results the values $N=4684.1$, $\alpha=16.87$, and $\beta=2.628$ \cite{Kniehl:2008zza} for the fragmentation function parameters.
\begin{figure}
	\begin{center}
		\includegraphics[width=0.7\linewidth,bb=137 42 690 690]{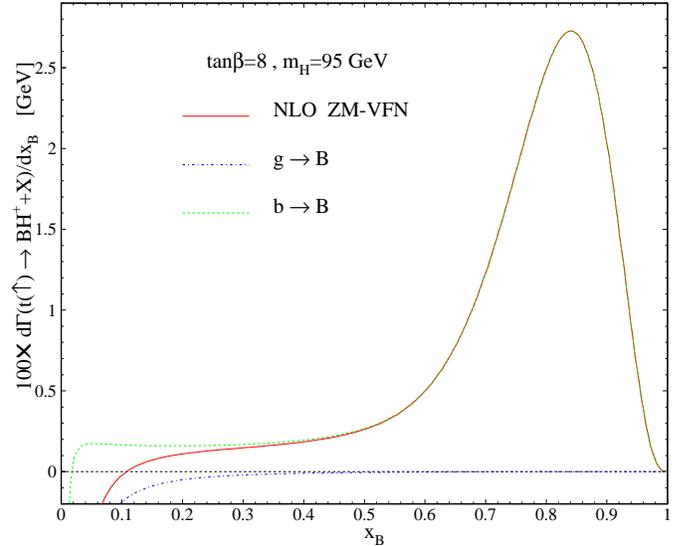}
		\caption{\label{fig0}%
				$d\Gamma/dx_B$ as a function of $x_B$ in the 2HDM with $m_{H^+}=95$~GeV and $\tan\beta=8$. The NLO result (solid line)  is broken up into the contributions due to $b\to B$ (dashed line) and $g\to B$ (dot-dashed line) fragmentation.}
	\end{center}
\end{figure}
\begin{figure}
	\begin{center}
		\includegraphics[width=0.7\linewidth,bb=137 42 690 690]{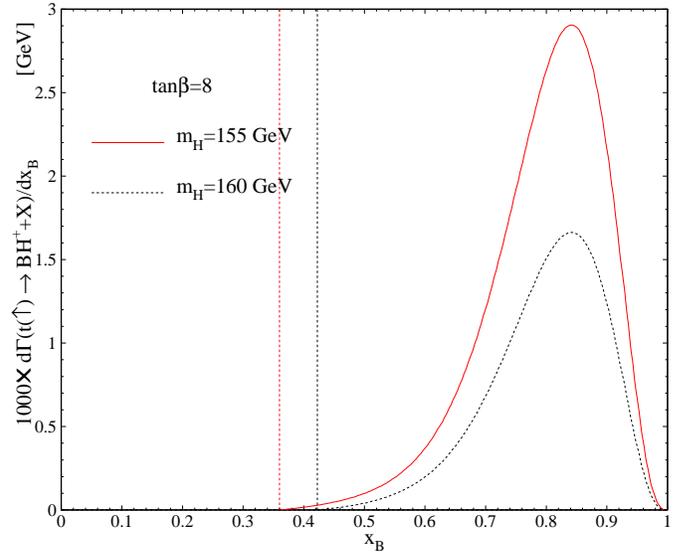}
		\caption{\label{fig1}%
			$x_B$ spectrum  in polarized top decay in the 2HDM  with $\tan\beta=8$ and  $m_{H^+}=155$ and $160$~GeV. Thresholds at $x_B$ are also shown.  Detail are discussed in the text.}
	\end{center}
\end{figure}
\begin{figure}
	\begin{center}
		\includegraphics[width=0.7\linewidth,bb=137 42 690 690]{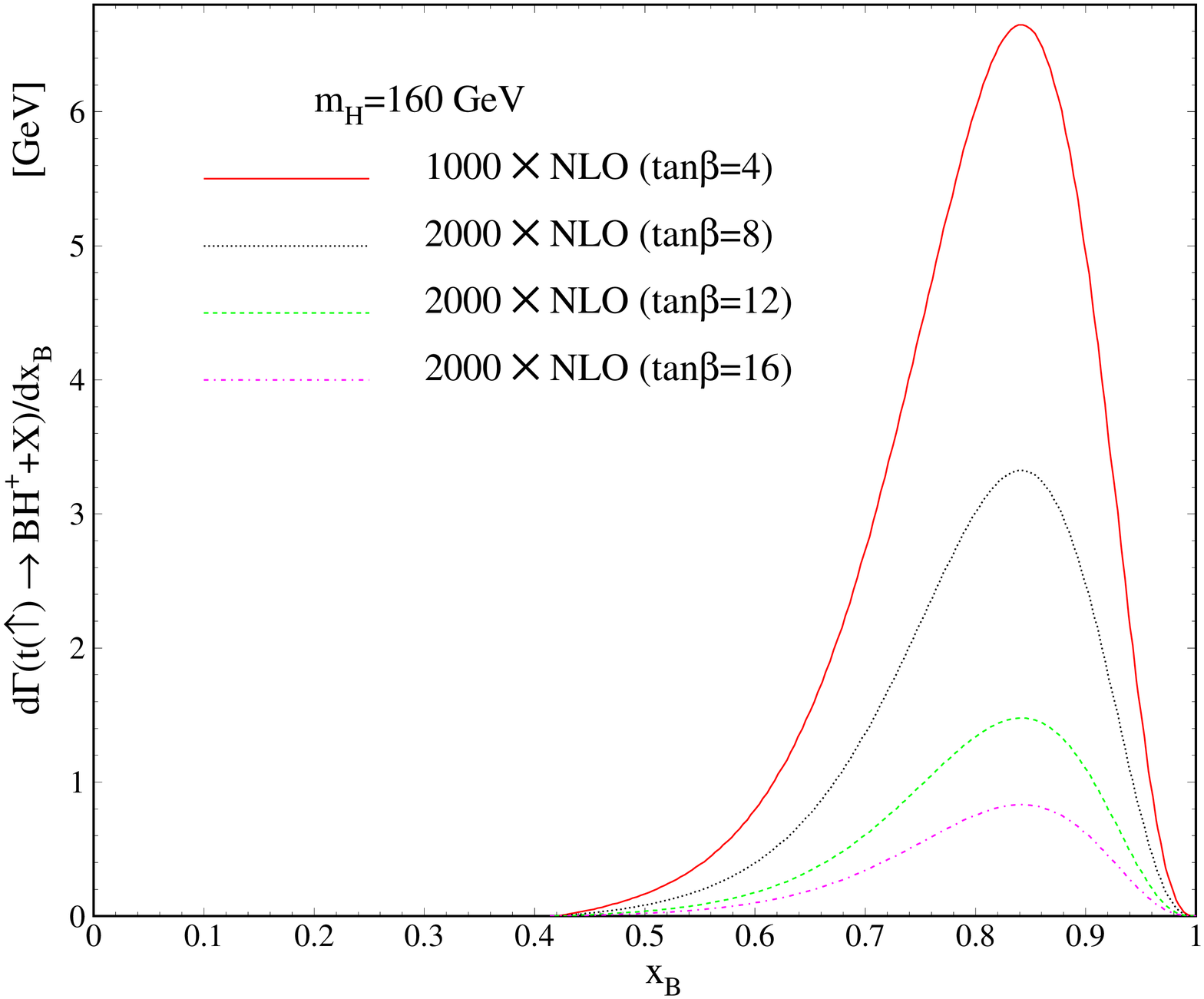}
		\caption{\label{fig2}%
			$x_B$ spectrum  in polarized top decay in the 2HDM $(t(\uparrow)\to BH^++X)$ with different values of $\tan\beta=4$, $8$, $12$ and $16$. The charged Higgs boson mass is fixed to $m_{H^+}=160$~GeV. For the large values of $\tan\beta$  the decay rate becomes  small.}
	\end{center}
\end{figure}
\begin{figure}
	\begin{center}
		\includegraphics[width=0.7\linewidth,bb=137 42 690 690]{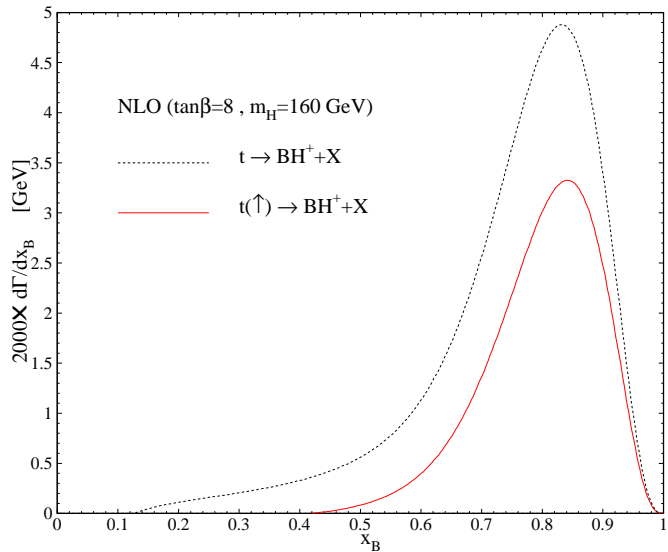}
		\caption{\label{fig3}%
			$d\Gamma/dx_B$ as a function of $x_B$ in the 2HDM considering the ZM-VFN ($m_b=0$) scheme. The polarized (solid line) and unpolarized (dashed line) partial decay rates are compared at NLO. We fixed $m_{H^+}=160$~GeV and $\tan\beta=8$. We also set $\mu_R=\mu_F=m_t$.}
	\end{center}
\end{figure}

Considering Fig.~\ref{plot}, where the charged Higgs masses $90\leq m_{H^+}\leq 100$~GeV (with $6<\tan\beta <10$) and  $140\leq m_{H^+}\leq 160$~GeV (with $3<\tan\beta<21$) are still unexcluded and could be possible masses, here, we study the scaled-energy spectrum of the B-hadron produced in the polarized top decay in the 2HDM. For this study we consider the  distribution $d\Gamma(t(\uparrow)\to BH^++X)/dx_B$ in the ZM-VFN scheme. 

In Fig.~\ref{fig0}, we show our prediction for the size of $d\Gamma/dx_B$, by considering the NLO  result (solid line) and the relative importance of the $b\to B$ (dashed line) and $g\to B$ (dot-dashed line) fragmentation channels at NLO, taking $\tan\beta=8$ and $m_{H^+}=95$~GeV. As is seen, the gluon fragmentation leads to an appreciable reduction in decay rate at low-$x_B$ region, for $x_B<0.35$. For example, the gluon splitting decreases the size of decay rate up to $45\%$ at $x_B=0.14$. For higher values of $x_B$, the $b\to B$ contribution is absolutely dominant.

In Fig.~\ref{fig1}, we show our prediction for the
size of the NLO corrections for $m_{H^+}=155$ (solid line) and $m_{H^+}=160$~GeV (dashed line) where $\tan\beta$ is set to $\tan\beta=8$. Here, the B-hadron mass creates a
threshold, e.g. at $x_B=2m_B/(m_t(1-y))\approx 0.42$ for $m_{H^+}=160$~GeV. As is seen, when $m_{H^+}$  increases the size of decay rate decreases but the peak position is approximately constant and independent of the charged Higgs mass.

Considering the unexcluded region from Fig.~\ref{plot} where $3\leq \tan\beta\leq 21$ is  allowed for $m_{H^+}=160$~GeV, in Fig.~\ref{fig2}  we study the energy spectrum of the B-hadron for 
different values of the $\tan\beta=4$, $8$, $12$ and $16$, where the mass of Higgs boson is fixed to $m_{H^+}=160$~GeV. As is seen when $\tan\beta$ increases the decay rate  decreases, as $\Gamma_0$ (\ref{gammaaa}) is proportional to $\cot^2\beta$. 

In Fig.~\ref{fig3}, the NLO energy spectrum of B-hadrons from the unpolarized  top decays $t\rightarrow BH^++X$ (dashed line) and the polarized ones $t(\uparrow)\rightarrow BH^++X$ (solid line) are compared considering $\tan\beta=8$ and $m_{H^+}=160$ GeV. 
Our results show that in these two cases the NLO corrections are similar in shape, however, the unpolarized distribution shows an more enhancement in size at NLO.

Our formalism elaborated here can be also extended  to the production of hadron species other than bottom-flavored hadrons, such as pions, kaons and protons, etc.,  using the nonperturbative $(b, g)\rightarrow \pi/K/P$ FFs extracted in our recent works \cite{Soleymaninia:2013cxa,Nejad:2015fdh},
relying on their universality and scaling violations.

\section{Conclusions}
\label{sec:three}

The top quark  is the heaviest elementary particle so that its large mass is a reason to rapid decay and, therefore, it has no time to
hadronize. Thus, it remains its full polarization content when it decays.
Due to $|V_{tb}|\approx 1$ of the CKM matrix, top quark decays are completely dominated by the mode $t\rightarrow W^++b$  within the SM   to a very high accuracy and in the theories beyond the SM including the two-Higgs-doublet,  the decay mode of light charged Higgses ($m_{H^\pm}<m_t$) is occurred  via $t\rightarrow H^++b$. The charged-Higgs bosons have been searched for in high energy experiments, in particular, at LEP and the Tevatron but they have  not been seen so far. But further searches are in progress so their discovery would indicate a signal of new physics beyond the SM. Among other things, the CERN LHC is a great top factory, producing around 90 million top pairs per year of running at design c.m. energy of 14 TeV. The existing and updating data will allow us to search for the charged Higgs boson if also the theoretical description and simulations are of proportionate quality.  

Since, bottom quarks produced through top decays  hadronize before they decay, then each $b$-jet  includes  a bottom flavored hadron
which, most of the times, is a B-meson. These mesons are identified by a displaced decay vertex associated which charged lepton tracks.\\
At LHC, the decay process $t\rightarrow BH^++X$ is proposed  to search for the light charged Higgs bosons  and evaluating the distribution in the scaled-energy ($x_B$) of B-mesons  in the top quark rest frame would be  of particular interest. For this study, one needs to evaluate the quantity $d\Gamma/dx_B$. The comparison of  future measurements of $d\Gamma/dx_B$ at the LHC with our NLO predictions   will be important for future tests of the Higgs coupling in the minimal supersymmetric SM (MSSM).\\   
In the present work,  using the ZM-VFN scheme we studied the $x_B$-distribution of B-meson in  the decay mode $t(\uparrow)\rightarrow BH^++X$ at NLO by working  on the type-I 2HDM scenario or a supersymmetric version of type-II;  MSSM.\\
In order to make our predictions  we, first, calculated an analytic
expression for  the NLO radiative corrections to the differential   decay width $d\Gamma(t(\uparrow)\to bH^+(+g))/dx_a (a=b, g)$ and then
employed the nonperturbative $(b,g)\to B$ FFs, relying on their universality and scaling violations \cite{collins}.
For our numerical analysis, considering the recent results reported by the CMS \cite{CMS:2014cdp} and ATLAS \cite{TheATLAScollaboration:2013wia} collaborations we restricted ourselves to the unexcluded regions of the  $m_{H^+}-\tan\beta$ parameter space which include 
$90\leq m_{H^+}\leq 100$~GeV (with $6<\tan\beta <10$) and  $140\leq m_{H^+}\leq 160$~GeV (with $3<\tan\beta<21$), see Fig.~\ref{plot}.

The top quark polarization is studied by the angular correlations between the top quark spin and its decay products momenta, so these spin-momenta correlations will allow the detailed studies of the top decay mechanism in the 2HDM. In our previous work \cite{MoosaviNejad:2011yp},   we studied the energy spectrum of B-meson in the 2HDM for the unpolarized decay mode. Here, we also compared the energy spectrum of B-mesons produced both through the unpolarized and polarized top decays. Results show a considerable difference between two distributions, however, they depend on the charged Higgs mass and $\tan\beta$.\\
Our formalism  can be also applied  for the production of other hadrons  such as pions, kaons and protons, etc.,  using the nonperturbative $(b, g)\rightarrow \pi/K/P$ FFs presented in  \cite{Soleymaninia:2013cxa,Nejad:2015fdh}.


\section{Acknowledgments}
\label{sec7}
We would like to thank Rebeca Gonzalez Suarez from the LHC top working group for reading the manuscript and also for importance discussion and comments. We warmly acknowledge M. Hashemi for valuable discussions and critical remarks.
We would also like to thank the CERN TH-PH division for its hospitality where a portion of this work was performed.

\end{document}